\documentclass[eqsecnum,nofootinbib,11pt]{revtex4}
\usepackage{amsmath}
   \usepackage{bm}
\usepackage{amsthm,amscd}
\usepackage{mathrsfs}
\usepackage{verbatim}
\usepackage{amsfonts,amsmath,latexsym,amssymb,bm}
\usepackage[american]{babel}
\usepackage{dsfont}
\usepackage[dvips]{graphicx}

\newcommand\bea{\begin{eqnarray}}
\newcommand\eea{\end{eqnarray}}
\newcommand{\C}{\mbox{${\rm I\!\!\!\!C }$}}
\newcommand{\beq}{\begin{eqnarray}}
\newcommand{\eeq}{\end{eqnarray}}
\newcommand{\nn}{\nonumber}

\begin{document}
\title{Pistons modeled by potentials}
\author{
Guglielmo Fucci\footnote{Electronic address: Guglielmo\textunderscore Fucci@Baylor.edu}, Klaus Kirsten\footnote{Electronic address: Klaus\textunderscore Kirsten@Baylor.edu}
and Pedro Morales\footnote{Electronic address: Pedro\textunderscore Morales@Baylor.edu}
\thanks{Electronic address: gfucci@nmt.edu}}
\affiliation{Department of Mathematics, Baylor University, Waco, TX 76798 USA
}
\date{\today}
\vspace{2cm}
\begin{abstract}

In this article we consider a piston modelled by a potential in the presence of extra dimensions. We analyze the functional determinant and the Casimir effect for this configuration. In order to compute the determinant and Casimir force we employ the zeta function scheme. Essentially, the computation reduces to the analysis of the zeta function associated with a scalar field living on an interval $[0,L]$ in a background potential. Although, as a model for a piston, it seems reasonable to assume a potential having compact support within $[0,L]$, we provide a formalism that can be applied to any sufficiently smooth potential.

\end{abstract}
\maketitle

\section{Introduction}
\label{sec:1}
In recent years piston configurations have received a surging interest in the Casimir effect community. The main reason for this fact is that pistons allow for an unambiguous prediction of forces which turn out to be divergence free \cite{cava04-69-065015}. The piston is usually represented by an infinitely thin movable plate at which the field has to satisfy some ideal boundary conditions. Different boundary conditions and various shapes of cross-sections have been analyzed and, as expected, the force heavily depends on the different possible choices, see, e.g., \cite{eliz09-79-065023,hert05-95-250402,kirs09-79-065019,mara07-75-085019,milt09-80-125028,teo09-819-431}. It is the aim of this article to represent pistons of finite thickness by compactly supported potentials. Physical properties of the pistons are encoded in the spectrum of the ordinary differential operator \begin{eqnarray} P := - \frac{d^2} {dx^2} + V(x), \label{1emi}\end{eqnarray} where $x\in I=[0,L]$ and $V(x)$ models the piston. The points $x=0$ and $x=L$ represent the positions of the fixed plates with suitable boundary conditions chosen. The piston at position $a$ is represented by the potential $V(x)$ which has a support strongly concentrated around $a$.

It is this problem to which an analysis in the space $I\times {\cal N}$ reduces after separation of variables, where ${\cal N}$ describes the cross section of the piston and additional dimensions, represented by a smooth Riemannian manifold possibly with boundary.

Although ultimately, in the context of pistons, our interest is in potentials with compact support within $I$, the formalism is developed for general potential $V(x)$.
We will use the zeta function scheme to evaluate various quantities of interest; for introductions to spectral zeta functions and its applications in physics see \cite{eliz95b,eliz94b,kirs02b}.
The relevant zeta function is represented as a contour integral where the boundary value of the unique solution to a given initial value problem enters. This representation is briefly described in Section \ref{sec:2} and subsequently used to evaluate the functional determinant, the Casimir energy and force. In order to analyze these quantities, the analytic continuation of the zeta function needs to be constructed. This will entail the knowledge of a certain asymptotic behavior of solutions to initial value problems, which is obtained in Section \ref{sec:3} using standard WKB techniques \cite{bend10b,mill06b}. The representation obtained can be used to find the functional determinant, Section \ref{sec:4}, and the Casimir energy and force, Section \ref{sec:5}. In Section \ref{sec:6} we restrict to compactly supported potentials and describe the resulting simplifications. In the Conclusions we summarize the most important findings and outline possible further applications of our approach.

\section{Contour representation of the piston zeta function}
\label{sec:2}
Let $M=[0,L]\times {\cal N}$, where ${\cal N}$ represents the cross section of the piston and the additional Kaluza-Klein dimensions. For simplicity we assume Dirichlet plates at $x=0$ and $x=L$ and we assume a sufficiently smooth potential $V(x)$ depending only on $x\in [0,L]$. With $y\in {\cal N}$, the relevant energy eigenvalues for a scalar field are determined by the second order differential operator \begin{eqnarray} L = - \frac{\partial^2}{\partial x^2} - \Delta_{{\cal N}} + V(x), \label{2emi}\end{eqnarray} together with Dirichlet boundary conditions at $x=0$ and $x=L$ and unspecified boundary conditions at the boundary of ${\cal N}$. Using separation of variables we write eigenfunctions in the form $$\phi (x,y) = X(x) \varphi (y),$$ where the $\varphi (y)$ are assumed as eigenfunctions of the Laplacian on ${\cal N}$, \begin{eqnarray} - \Delta _{{\cal N}} \varphi _\ell (y) = \eta_\ell ^2 \varphi _\ell (y). \label{3emi} \end{eqnarray} This implies that the eigenvalues $\lambda$ of $L$ are given as $$\lambda = \nu^2 + \eta_\ell ^2,$$ where $\nu^2$ is determined by \begin{eqnarray} \left( - \frac{\partial^2} {\partial x^2} + V(x) \right) X_\nu (x) = \nu^2 X_\nu (x), \quad \quad X_\nu (0) = X_\nu (L) =0.\label{4emi}\end{eqnarray}
Formally, the zeta function of $L$ can therefore be written as \begin{eqnarray} \zeta (s) = \sum_{\ell, \nu} (\nu^2 + \eta_\ell^2)^{-s} \quad \quad \mbox{for } \Re s > \frac D 2 , \label{5emi}\end{eqnarray} with $D=\mbox{dim} (M)$, the dimension of $M$. Note that without specifying ${\cal N}$ the spectrum $\eta _\ell$ is not known, and also $\nu^2$ will not be known unless $V(x)$ is one of the very few potentials allowing for a closed solution of eq. (\ref{4emi}).

Despite this lack of knowledge an analytical continuation of $\zeta (s)$ in eq. (\ref{5emi}) can be constructed and properties of $\zeta (s)$ on $M$ can be given in terms of the zeta function of ${\cal N}$ defined by
$$ \zeta_{{\cal N}} (s) = \sum_\ell \eta_\ell ^{-2s}  \quad \quad \mbox{for } \Re s > \frac{D-1} 2 .$$ The strategy for the analysis of $\zeta (s)$ in eq. (\ref{5emi}) is to rewrite the series as a contour integral using the argument principle or Cauchy's residue theorem \cite{conw78b}. Instead of considering the eigenvalue problem in eq. (\ref{4emi}) we consider the {\it initial value problem} \cite{kirs03-308-502,kirs04-37-4649} $$ \left( - \frac{\partial^2}{\partial x^2} + V(x) \right) u_\mu (x) = \mu^2 u_\mu (x), \quad \quad u_\mu (0) =0, \quad u_\mu ' (0) =1, $$ where $\mu \in {\C}$. The eigenvalues $\nu^2$ of the original problem are recovered as solutions to the secular equation \beq u_\mu (L) =0; \label{6emi} \eeq note, that $u_\mu (L)$ is an analytic function of $\mu$. Eq. (\ref{5emi}), for $\Re s > D/2$, can then be rewritten as $$ \zeta (s) = \frac 1 {2\pi  i} \sum_\ell \int\limits_\gamma d\mu (\mu^2 + \eta_\ell ^2)^{-s} \frac d {d\mu} \ln u_\mu (L), $$ where $\gamma$ encloses all solutions to eq. (\ref{6emi}), which are assumed to be on the positive real axis; the changes necessary when zero modes or finitely many negative eigenvalues are present are given in \cite{kirs04-37-4649}.

Let us next consider the contributions from each $\ell$ by analyzing $$ \zeta _\ell (s) = \frac 1 {2\pi i} \int\limits_\gamma d\mu (\mu^2 + \eta_\ell ^2)^{-s} \frac d {d\mu} \ln u_\mu (L).$$ Deforming the contour, as usual, to the imaginary axis we find
 \beq \zeta _\ell (s) = \frac{\sin \pi s} \pi \int\limits_{\eta_\ell} ^\infty dk \, (k^2 - \eta_\ell ^2)^{-s} \frac d {dk} \ln u_{ik} (L), \label{7emi} \eeq valid for $1/2< \Re s < 1.$ In order to construct a representation of $\zeta _\ell (s)$ that is valid in a region $\Re s < 1/2$, as is needed for the functional determinant and the Casimir energy, we add and subtract the large-$k$ asymptotics of $u_{ik} (L)$.
\section{Asymptotic behavior of boundary values for differential equations}
\label{sec:3}
The next mathematical task therefore is to determine the large-$k$ asymptotics of the unique solution to the initial value problem
\beq \left( - \frac{d^2} {dx^2} + V(x) + k^2 \right) u_{ik} (x) =0, \quad \quad u_{ik} (0) =0, \quad u_{ik} ' (0) =1.\label{8emi} \eeq
We note that the differential equation in (\ref{8emi}) has exponentially growing and exponentially decaying terms \cite{bend10b,mill06b}. Although, ultimately, for (\ref{7emi}) we will only need the exponentially growing part, we first have to consider linear combinations of the two in order to be able to impose the initial conditions in eq. (\ref{8emi}).

It is convenient and standard to introduce $$S(x,k) = \partial _x \ln \psi_k (x),$$ where $\psi_k (x)$ satisfies
\beq \left( - \frac {d^2}{dx^2} + V(x) + k^2 \right) \psi _k (x) =0. \label{9aemi} \eeq
The differential equation satisfied by $S(x,k)$ turns out to be \beq  S ' (x,k) = k^2 + V(x) - S^2 (x,k) , \label{9emi}\eeq
where the prime indicates differentiation with respect to $x$. As $k\to\infty$, the function $S(x,k)$ can be seen to have the asymptotic form $$S(x,k) = \sum _{i=-1}^\infty k^{-i} S_i (x), $$ where the asymptotic orders $S_i (x)$ are given by \beq S_{-1}(x) &=& \pm 1, \quad S_0 (x) =0, \quad S_1 (x) = \pm \frac{V(x)} 2 , \label{10emi}\\
S_{i+1} (x) &=& \mp \frac 1 2 \left(S_i ' (x) + \sum_{j=0}^i S_j (x) S_{i-j} (x) \right) .\nn\eeq
It is clear that an arbitrary number of asymptotic orders can be easily evaluated using an algebraic computer program. The two different signs in (\ref{10emi}) produce the indicated exponentially growing and decaying solutions $\psi_k (x)$ of (\ref{9aemi}). We denote solutions of (\ref{9emi}) corresponding to these two signs by $S^+ (x,k)$ and $S^- (x,k)$. The associated solutions of (\ref{9aemi}) then have the form $$ \psi_k ^\pm (x) = A^\pm \exp \left\{ \int\limits_0^x dt \,\, S^\pm (t,k) \right\}.$$ The large-$k$ behavior for $u_{ik} (x)$ is obtained by considering the linear combination
$$ u_{ik} (x) = A^+ \exp \left\{ \int\limits_0^x dt \,\, S^+ (t,k) \right\} + A^- \exp \left\{ \int\limits_0^x dt \,\, S^- (t,k) \right\}, $$ together with the initial conditions in (\ref{8emi}) still to be imposed. These initial conditions imply $$A^+ = - A^-, \quad \quad A^+ = \frac 1 {S^+ (0,k) - S^- (0,k)};$$ note, that without including the $S^- (x,k)$ part the initial conditions could not be satisfied.

We are now in the position to write out the large-$k$ behavior for $u_{ik} (L)$. Let $E(k)$ denote exponentially damped terms as $k\to \infty$. First, we see that, as $k\to\infty$, $$u_{ik} (L) = \frac 1 {S^+ (0,k) - S^- (0,k)} \exp \left\{ \int\limits_0^L dt \,\, S^+(t,k) \right\} + E(k), $$ and therefore \beq \ln u_{ik} (L) &=& - \ln \left( S^+ (0,k) - S^- (0,k)\right) + \int\limits_0^L dt \,\, S^+(t,k) + E(k) \nn\\
&=& - \ln (2k) + k L+ \sum_{j=0}^\infty d_j k^{-j} + E(k) , \nn\eeq where the $d_j$'s are easily determined from eq. (\ref{10emi}).
For instance, the first six are given explicitly by
\beq d_0&=&0,\quad\quad d_1=\frac{1}{2}\int\limits_0^L dt\,\,V(t),\quad\quad d_2 = - \frac 1 4 [V(L) + V(0) ] ,\nn\\
d_3&=&\frac{1}{8}[V'(L)-V'(0)]-
\frac{1}{8}\int\limits_0^Ldt\,\, V^2(t),\nn\\ d_4&=&-\frac{1}{16}[V''(L)+V''(0)]+\frac{1}{8}[V^2(L)-V^2(0)],\label{dsubi}\nn\\
d_5&=&\frac{1}{32}[V^{(3)}(L)-V^{(3)}(0)]-\frac{5}{32}[V(L)V'(L)-V(0)V'(0)]+\frac{1}{16}
\int\limits_0^Ldt\,\,V^3(t)\nn\\
&-&\frac{1}{32}\int\limits_0^Ldt\,\,V(t)V''(t).\eeq
In what follows, the potential is always assumed to be as smooth as necessary for the asymptotic orders given by these formulas, and higher ones if needed, to be well defined.

Subtracting and adding the asymptotic behavior up to the order $k^{-N}$, the zeta function naturally splits into two parts,
$$ \zeta_\ell (s) = \zeta _\ell ^{(f)} (s) + \zeta _\ell ^{(as)} (s), $$ where \beq & &\hspace{-.7cm}\zeta _\ell^{(f)} (s) = \frac{\sin \pi s} \pi \int\limits_{\eta_\ell} ^\infty dk \,\, (k^2 - \eta_\ell^2)^{-s} \frac d {dk} \left\{ \ln u_{ik} (L) -kL + \ln (2k) -\sum_{j=0}^N d_j k^{-j} \right\} , \label{11emi}\\
& &\hspace{-.7cm}\zeta_\ell ^{(as)} (s) =  \frac{\sin \pi s} \pi \int\limits_{\eta_\ell} ^\infty dk \,\, (k^2 - \eta_\ell^2)^{-s} \frac d {dk} \left\{ kL - \ln (2k) +\sum_{j=0}^N d_j k^{-j} \right\} . \label{12emi}\eeq
The $k$-integrals in $\zeta_\ell ^{(as)} (s)$ are easily done, yielding
\beq \zeta_\ell ^{(as)} (s) = \frac 1 {2 \Gamma (s)} \left\{ \frac{L \Gamma \left( s- \frac 1 2 \right)}{\sqrt \pi} \eta_\ell ^{1-2s}
- \Gamma (s) \eta_\ell ^{-2s} - \sum_{j=1}^N j d_j \frac{ \Gamma \left( s+ \frac j 2 \right)}{\Gamma \left( 1 + \frac j 2 \right)} \eta_\ell^{-j-2s} \right\}.\nn\eeq
After summing over $\ell$, the representation obtained is then valid for $1>\Re s > (D-N-2)/2$ and it reads
\beq & &\hspace{-.5cm}\zeta ^{(f)}(s) = \frac{\sin \pi s} \pi \sum_\ell \int\limits_{\eta _\ell}^\infty dk \,\, (k^2 - \eta_\ell^2)^{-s} \frac d {dk} \left\{ \ln u_{ik} (L) - kL + \ln (2k) -\sum_{j=0}^N \frac{d_j}{ k^{j}} \right\} , \label{13emi}\\
 & &\hspace{-.5cm}\zeta ^{(as)} (s) = \frac 1 {2 \Gamma (s)} \left\{  \frac{ L \Gamma \left( s- \frac 1 2 \right)}{\sqrt \pi} \zeta _{{\cal N}} \left( s- \frac 1 2 \right)
- \Gamma (s) \zeta _{{\cal N}} (s) \right.\nn\\
& &\hspace{3.0cm}\left.- \sum_{j=1}^N j d_j \frac{ \Gamma \left( s+ \frac j 2 \right)}{\Gamma \left( 1 + \frac j 2 \right)} \zeta_{{\cal N}} \left( s + \frac j 2\right) \right\}.\label{14emi}\eeq
In particular, choosing $N=D-1$, respectively $N=D$, the representation can be used to compute the determinant, respectively the Casimir energy. This will be done in the next sections.
\section{Functional determinants}
\label{sec:4}
In this section we will evaluate the functional determinant, or, equivalently, $\zeta ' (0)$, using the representation of $\zeta(s)$ given by eqs. (\ref{13emi}) and (\ref{14emi}). The contribution from $\zeta ^{(f)}(s)$ is trivially obtained and it reads $$ {\zeta ^{(f)}} ' (0) = - \sum_\ell \left[ \ln u_{i\eta _\ell} (L) - L \eta _\ell + \ln (2\eta_\ell) - \sum_{ j=1}^{D-1} d_j \eta_\ell^{-j}\right],$$ as the sum over $\ell$ converges by construction. For the evaluation of the contribution from $\zeta ^{(as)} (s)$, let us note that in the general situation considered, namely ${\cal N}$ can be a manifold of any dimension with or without boundary, for $j=1,...,(D-1)$, we have the following Laurent expansion \beq \zeta_{{\cal N}} \left( \frac j 2 + \epsilon\right) = \frac 1 {\epsilon} \mbox{Res } \zeta_{{\cal N}} (j/2) + \mbox{FP } \zeta _{{\cal N}} (j/2) + {\cal O} (\epsilon ).\label{mero}\eeq Using standard properties of $\Gamma$-functions, this immediately gives \beq {\zeta ^{(as)}} ' (0) &=& - L\cdot \mbox{FP } \zeta_{{\cal N}} \left( -\frac 1 2\right) + L\cdot \mbox{Res } \zeta_{{\cal N}} \left( -\frac 1 2 \right) (-2 + \ln 4) - \frac 1 2 \zeta_{{\cal N}} ' (0) \nn\\
& &- \sum_{j=1}^{D-1} d_j \left( \mbox{FP } \zeta_{{\cal N}} \left( \frac j 2 \right) + \mbox{Res } \zeta_{{\cal N}} \left( \frac j 2 \right) \left( \gamma + \psi \left( \frac j 2 \right) \right)\right) .\nn\eeq
Adding up these two pieces gives the answer for the functional determinant on $M$ in terms of the spectral zeta function on ${\cal N}$. Once the 'base' manifold ${\cal N}$ is specified, more explicit results can be given.
\section{Casimir energy}
\label{sec:5}
For the Casimir energy we set $N=D$ and use again the Laurent series (\ref{mero}) for $\zeta_{{\cal N}} (j/2+\epsilon )$. In this case, we find
\beq \zeta ^{(f)} (-1/2) &=& - \frac 1 \pi \sum_\ell \int\limits_{\eta_\ell } ^\infty dk \,\, (k^2 - \eta_\ell^2)^{1/2} \frac d {dk} \left\{ \ln u_{ik} (L) - kL + \ln (2k) - \sum_{j=1}^D \frac{d_j}{ k^{j}} \right\} , \nn
\eeq
\beq
\zeta ^{(as)} (-1/2+\epsilon) &=& \frac 1 \epsilon \left\{ \frac L {4\pi} \zeta_{{\cal N}} (-1) - \frac 1 2 \mbox{Res } \zeta_{{\cal N}} (-1/2) + \frac{d_1 \zeta_{{\cal N}} (0)} {2\pi} \right.\nn\\
& &\hspace{2.0cm}\left. + \sum_{j=2}^D \frac{d_j}{2\sqrt \pi} \frac{\Gamma \left( \frac{j-1} 2 \right)} {\Gamma \left( \frac j 2 \right)} \mbox{Res } \zeta_{{\cal N}} \left( \frac{j-1} 2 \right) \right\}\nn\\
& &\hspace{-2.70cm}+ \frac L {4\pi} \left( \zeta_{{\cal N}} ' (-1) + \zeta_{{\cal N}} (-1) \left( -1 + \ln 4\right) \right)- \frac 1 2 \mbox{FP } \zeta_{{\cal N}} (-1/2) \nn\\
& &\hspace{-2.7cm}+\frac{d_1}{2\pi} \left( \zeta_{{\cal N}}  ' (0) + \zeta_{{\cal N}} (0) \left( -2 + \ln 4\right)\right) \nn\\
& &\hspace{-2.7cm}+\sum_{j=2}^D \frac{ d_j} {2\sqrt \pi} \frac{\Gamma \left( \frac{j-1} 2 \right)} {\Gamma \left( \frac j 2 \right) } \left( \mbox{FP  } \zeta_{{\cal N}} \left( \frac{j-1} 2 \right) + \mbox{Res } \zeta_{{\cal N}} \left( \frac{j-1} 2 \right) \left[ H_{(k-3)/2} - 2 + \ln 4\right] \right), \nn\eeq
with the harmonic numbers $$H_n = \sum_{i=1}^n \frac 1 i.$$
Multiplying by $1/2$ and adding up the two pieces, the Casimir energy follows. Again, when ${\cal N}$ is specified more explicit results can be given. In general, the Casimir energy is divergent, but as seen below, the resulting forces on pistons are finite.
\section{Compactly supported potentials}
\label{sec:6}
In order to reasonably talk about the force on a piston modeled by a potential we now assume the potential
$V(x)$ to have compact support within the interval $[0,L]$; namely, we assume that it does not vanish for $x\in[a-\epsilon,a+\epsilon ]\subset[0,L]$. This can be considered a model for a piston of thickness $2\epsilon$.
In this case the asymptotic behavior of $u_{ik} (L)$ will be independent of $a$ as the integrals over $V(x)$
and its powers and derivatives are independent of $a$ (as long as the support is within the interval $[0,L]$).
This can be seen explicitly in (\ref{dsubi}). It is here that sufficiently smooth potentials become a necessary assumption in order for these formulas, and the corresponding ones for higher orders, to be well defined. These formulas also simplify further because $V(0)= V(L) =0$. The corresponding result can be used to write down the Casimir energy. Because of the independence of the asymptotic behavior of $u_{ik} (L)$ from $a$, in these circumstances we immediately obtain \beq F_{Cas} = - \frac 1 2 \frac{\partial} {\partial a}
\zeta \left( - \frac 1 2 \right) = \frac 1 {2\pi } \sum_\ell \int\limits_{\eta _\ell} ^\infty dk \,\,
(k^2 - \eta_\ell^2)^{1/2} \frac \partial {\partial a} \frac \partial {\partial k} \ln u_{ik} (L).\nn\eeq
The force, in particular its sign, is encoded in boundary values of an initial value problem to an ordinary differential equation.

Results for the determinant are also easily written down from Section \ref{sec:4} with the simpler $d_j$'s used.
\section{Conclusions and outlook}
\label{sec:7}
In this article we have provided a formalism that allows for the evaluation of functional determinants and Casimir energies and forces for the configuration of a generalized piston. Results are as explicit as they can be without specifying the cross section and the additional Kaluza-Klein dimensions that might be present.

With these results available, one can obtain very explicit answers for a given cross section of the piston and specified Kaluza-Klein dimensions. Furthermore one can use the potential to mimic material properties of the piston. For all cases, with very few exceptions for particular potentials, a numerical determination of the boundary value $u_{ik} (L)$ will be necessary.

Along the same lines different boundary conditions at $x=0$ and $x=L$ can be considered.

Work along these lines is currently in progress.\\[.5cm]

{\bf Acknowledgement:} KK is supported by National Science Foundation grant PHY--0554849.


\end{document}